\documentclass[12pt]{article}
\usepackage{graphicx}
\def\bra#1{\left\langle #1\right|}
\def\ket#1{\left| #1\right\rangle}
\newcommand{\bers}{\begin{eqnarray*}}
\newcommand{\eers}{\end{eqnarray*}}
\newcommand{\bt}{\begin{itemize}}
\newcommand{\et}{\end{itemize}}
\def\beq{\begin{equation}}
\def\eeq{\end{equation}}
\def\bea{\begin{eqnarray}}
\def\eea{\end{eqnarray}}
\def\nn{\nonumber}
\def\M{m_{\Lambda_b}}
\def\m{m_{\Lambda_c}}
\def\mv{m_{V}}
\def\lb{\Lambda_b}
\def\lc{\Lambda_c}
\def\sla#1{\raise.15ex\hbox{$/$}\kern-.57em #1}

\def\sss{\scriptscriptstyle}
\def\bd{B_d^0}
\def\bdbar{{\overline{B_d^0}}}

\def\barp{{\raise.35ex\hbox
{${\sss (}$}}---{\raise.35ex\hbox{${\sss )}$}}}
\def\bdbarp{\hbox{$B_d$\kern-1.4em\raise1.4ex\hbox{\barp}}}
\def\bsbarp{\hbox{$B_s$\kern-1.4em\raise1.4ex\hbox{\barp}}}

\def\roughly#1{\mathrel{\raise.3ex\hbox
{$#1$\kern-.75em\lower1ex\hbox{$\sim$}}}}

\def\npb#1#2#3{{\it Nucl.\ Phys.} {\bf B#1} (#2) #3}
\def\plb#1#2#3{{\it Phys.\ Lett.} {\bf #1B} (#2) #3}

\def\zpc#1#2#3{{\it Zeit.\ Phys.} {\bf C#1} (#2) #3}

\newread\epsffilein 
\newif\ifepsffileok 
\newif\ifepsfbbfound 
\newif\ifepsfverbose 
\newdimen\epsfxsize 
\newdimen\epsfysize 
\newdimen\epsftsize 
\newdimen\epsfrsize 
\newdimen\epsftmp 
\newdimen\pspoints 
\def\epsfbox#1{\global\def\epsfllx{72}\global\def\epsflly{72}%
 \global\def\epsfurx{540}\global\def\epsfury{720}%
 \def\lbracket{[}\def\testit{#1}\ifx\testit\lbracket
 \let\next=\epsfgetlitbb\else\let\next=\epsfnormal\fi\next{#1}}%
\def\epsfgetlitbb#1#2 #3 #4 #5]#6{\epsfgrab #2 #3 #4 #5 .\\%
 \epsfsetgraph{#6}}%
\def\epsfnormal#1{\epsfgetbb{#1}\epsfsetgraph{#1}}%
\def\epsfgetbb#1{%
\openin\epsffilein=#1
\ifeof\epsffilein\errmessage{I couldn't open #1, will ignore it}\else
 {\epsffileoktrue \chardef\other=12
 \def\do##1{\catcode`##1=\other}\dospecials \catcode`\ =10
 \loop
 \read\epsffilein to \epsffileline
 \ifeof\epsffilein\epsffileokfalse\else
 \expandafter\epsfaux\epsffileline:. \\%
 \fi
 \ifepsffileok\repeat
 \ifepsfbbfound\else
 \ifepsfverbose\message{No bounding box comment in #1; using defaults}\fi\fi
 }\closein\epsffilein\fi}%
\def\epsfclipstring{}
\def\epsfsetgraph#1{%
 \epsfrsize=\epsfury\pspoints
 \advance\epsfrsize by-\epsflly\pspoints
 \epsftsize=\epsfurx\pspoints
 \advance\epsftsize by-\epsfllx\pspoints
 \epsfxsize\epsfsize\epsftsize\epsfrsize
 \ifnum\epsfxsize=0 \ifnum\epsfysize=0
 \epsfxsize=\epsftsize \epsfysize=\epsfrsize
 \epsfrsize=0pt
 \else\epsftmp=\epsftsize \divide\epsftmp\epsfrsize
 \epsfxsize=\epsfysize \multiply\epsfxsize\epsftmp
 \multiply\epsftmp\epsfrsize \advance\epsftsize-\epsftmp
 \epsftmp=\epsfysize
 \loop \advance\epsftsize\epsftsize \divide\epsftmp 2
 \ifnum\epsftmp>0
 \ifnum\epsftsize<\epsfrsize\else
 \advance\epsftsize-\epsfrsize \advance\epsfxsize\epsftmp \fi
 \repeat
 \epsfrsize=0pt
 \fi
 \else \ifnum\epsfysize=0
 \epsftmp=\epsfrsize \divide\epsftmp\epsftsize
 \epsfysize=\epsfxsize \multiply\epsfysize\epsftmp
 \multiply\epsftmp\epsftsize \advance\epsfrsize-\epsftmp
 \epsftmp=\epsfxsize
 \loop \advance\epsfrsize\epsfrsize \divide\epsftmp 2
 \ifnum\epsftmp>0
 \ifnum\epsfrsize<\epsftsize\else
 \advance\epsfrsize-\epsftsize \advance\epsfysize\epsftmp \fi
 \repeat
 \epsfrsize=0pt
 \else
 \epsfrsize=\epsfysize
 \fi
 \fi
 \ifepsfverbose\message{#1: width=\the\epsfxsize, height=\the\epsfysize}\fi
 \epsftmp=10\epsfxsize \divide\epsftmp\pspoints
 \vbox to\epsfysize{\vfil\hbox to\epsfxsize{%
 \ifnum\epsfrsize=0\relax
 \includegraphics{#1}%
 \else
 \epsfrsize=10\epsfysize \divide\epsfrsize\pspoints
 \includegraphics{#1}%
 \fi
 \hfil}}%
\global\epsfxsize=0pt\global\epsfysize=0pt}%
 {\catcode`\%=12 \global\let\epsfpercent=
\long\def\epsfaux#1#2:#3\\{\ifx#1\epsfpercent
 \def\testit{#2}\ifx\testit\epsfbblit
 \epsfgrab #3 . . . \\%
 \epsffileokfalse
 \global\epsfbbfoundtrue
 \fi\else\ifx#1\par\else\epsffileokfalse\fi\fi}%
\def\epsfempty{}%
\def\epsfgrab #1 #2 #3 #4 #5\\{%
\global\def\epsfllx{#1}\ifx\epsfllx\epsfempty
 \epsfgrab #2 #3 #4 #5 .\\\else
 \global\def\epsflly{#2}%
 \global\def\epsfurx{#3}\global\def\epsfury{#4}\fi}%
\def\epsfsize#1#2{\epsfxsize}
\pagestyle{plain}

\begin{document}

\vskip0.5truecm

\begin{center} 

{\large \bf
 Nonleptonic $\Lambda_b$ decays to
$D_s(2317)$, $D_s(2460)$ and other final states  
in Factorization. 
 }
\vspace*{1.0cm}
\vskip1cm
{\large Alakabha Datta\footnote{email: datta@physics.utoronto.ca}${}^{a}$,
Harry J. Lipkin\footnote{email: harry.lipkin@weizmann.ac.il}${}^{b}$
and Patrick J. O'Donnell
\footnote{email: odonnell@physics.utoronto.ca}${}^{a}$ 
\vskip0.3cm
${}^{a}${\it  Department of Physics, University of Toronto} \\
{\it 60 St George Street, Toronto, ON, Canada M5S 1A7}. \\
${}^{b)}$ {\it
Department of Particle Physics,\\
Weizmann Institute,\\
Rehovot 76100, Israel \\and\\
School of Physics and Astronomy, \\
Raymond and Beverly Sackler Faculty of Exact Sciences\\
Tel-Aviv University,\\
Tel-Aviv, Israel \\
and\\
High Energy Physics Division\\
 Argonne National Laboratory\\
Argonne, IL 60439-4815, USA\\
}}

\vskip0.5cm
\bigskip
(\today)
\vskip0.5cm
\newpage
{\Large Abstract}\\
\vskip3truemm
\parbox[t]{\textwidth} 
{We  consider nonleptonic Cabibbo--allowed  $\Lambda_b$ decays  in the
factorization approximation.   We calculate nonleptonic  decays of the
type $ \Lambda_b  \to \Lambda_c P $ and $ \Lambda_b  \to \Lambda_c V $
relative to $\bdbar \to D^+ P$ and $\bdbar \to D^+ V$ where we include
among the  pseudoscalar states(P) and  the vector states(V)  the newly
discovered $D_s$ resonances,  $D_s(2317)$ and $D_s(2460)$.
In the ratio of $\lb$ decays to $D_s(2317)$ and $D_s(2460)$
relative to the $\bdbar$ decays to these states, the poorly known 
decay constants of $D_s(2317)$ and $D_s(2460)$ cancel leading to
predictions that can shed light on the nature of these new states.
In general,
we  predict the  $\lb$  decays  to be  larger  than the  corresponding
$\bdbar$ decays and in particular we find the branching ratio for $\lb
\to \lc  D_s(2460)$ can  be between four  to five times  the branching
ratio  for  $\bdbar \to  D^+  D_s(2460)$.  This  enhancement of  $\lb$
branching  ratios follows primarily  from the  fact that  more partial
waves  contribute  in  $\lb$  decays  than in  $\bdbar$  decays.   Our
predictions are largely independent  of model calculations of hadronic
inputs like form factors and decay constants.}
\end{center}
\thispagestyle{empty} \newpage \setcounter{page}{1}
\textheight 23.0 true cm \baselineskip=14pt
\section{Introduction}

Nonleptonic  decays are widely  used to  obtain information  about the
elements of  the CKM matrix in  the Standard Model(SM), as  well as to
obtain   insights   about  the   non-perturbative   aspects  of   QCD.
Nonleptonic decays in the  $B$ and $\Lambda_b$ systems are interesting
since the  heavy mass of the $b$  quark relative to the  scale of soft
non-perturbative   physics  allows   for  simplifications   and  makes
tractable the difficult problem of calculating nonleptonic decays.

The  nonleptonic  decays  of  the  $\Lambda_b$  baryon  have  received
relatively  less  attention than  those  of  the  $B$ meson.   In  the
$\Lambda_b$ baryon, the spin of the baryon is carried by the $b$ quark
with the  light diquark in  a spin-- and isospin--singlet  state. This
fact  plays an  important  role in  $\Lambda_b$ decays\cite{dlo1}  and
leads to  simplification of the non-perturbative  dynamics involved in
these decays.   Because of this  spin correlation between the  the $b$
and the $\Lambda_b$ polarized $\Lambda_b$ decays can provide important
information   about   the   weak   interaction  of   the   $b$   quark
\cite{bdl1}. Nonleptonic  $\Lambda_b$ decays can therefore  be used to
test the SM and to obtain insights into non-perturbative QCD.

In this work we consider Cabibbo--allowed $\Lambda_b$ decays.
Cabibbo--allowed $\Lambda_b$ and $B$ decays are usually calculated
using factorization. We will also concentrate only on the factorizable
part and discuss briefly nonfactorizable effects later in this
section.

The factorizable amplitude  is expressed in terms of  form factors and
decay constants.  However  the form factors and, barring  a few cases,
the   decay  constants  are   unknown  hadronic   inputs.   Therefore,
predictions  for  nonleptonic  $\Lambda_b$  decays,  depend  on  model
calculation of  form factors and decay  constants and can  have a wide
range even within the factorization assumption \cite{lambdabold}.  Our
purpose in this paper is to obtain predictions for $\Lambda_b$ decays,
within  factorization,   using  the   heavy  $m_b$  limit   and  using
experimental inputs.

The method we use is the following: instead of directly calculating
the $\Lambda_b$ decays we consider instead the ratio of
Cabibbo--allowed $\Lambda_b$ decays relative to the corresponding
Cabibbo--allowed $B$ decays.  The branching ratios for the $\Lambda_b$
decays can then be obtained by simply using the experimental numbers
for the Cabibbo--allowed $B$ decays.  One obvious advantage of
considering such ratios is that the dependence on decay constants drop
out in the ratio.  Furthermore, in the heavy $m_b$ limit, these ratios
can be expressed as ratios of squared form factors. In the heavy $m_b$
limit all form factors can be related to one single form factor and a
dimensional constant representing the effective mass of the light
degrees of freedom in the $\lb$ baryon.  These ratios of form factors
are obtained using a mild assumption about the $q^2$ behavior of the
form factors and the measurement of ${BR[\Lambda_b \to \Lambda_c
  \pi^-] / BR[\bdbar \to D^+ \pi^-]}$. Our predictions turn out to be
minimally dependent on hadronic inputs like form factors and decay
constants.

Another advantage  of calculating ratios  of branching ratios  is that
some of the nonfactorizable amplitudes cancel in the ratio. To see how
this happens consider the decays $\lb  \to \lc P$ and $ \bdbar \to D^+
P$.  Now  the  underlying quark  transition  is  $  b  \to c  P$.  The
corrections to factorization can arise from gluon emmision between the
$b$ or the $c$ quark and  the quark constituents of $P$. However these
corrections are the  same for $\lb$ and $\bdbar$  decays and so cancel
in  the ratio  of  their amplitudes.   Gluon  emmisions involving  the
spectator quark  in $\bdbar$  and the spectator  diquark in  $\lb$ may
also be  similar given the fact  that the diquark in  $\lb$ belongs in
the  ${\overline  3}$ under  color  $SU(3)_c$  as  does the  spectator
anti-quark in  $\bdbar$ and  so both the  spectators may  have similar
color  interactions.   Furthermore,  within factorization,  the  small
perturbative corrections to the form factors will also cancel.

Nonleptonic decays involving the newly discovered $D_s(2317)$
\cite{babarDs} and $D_s(2460)$ \cite{cleoDs} states are of particular
interest. It was shown in Ref. \cite{do1} that nonleptonic $\bdbar$
decays involving these states can provide clues to the true nature of
these states which is still not known \cite{barneslipkin, jackson,
  bardeenhill,sandip}. It will be therefore interesting to see if
these new $D_s$ resonances show up in the $\lb$ decays and how the
rates for these decays compare to the $\bdbar$ decays involving the
new $D_s$ states. As shown in Ref. \cite{do1} nonleptonic $\bdbar$
decays to $D_s(2317)$ and $D_s(2460)$ involve the poorly known decay
constants of these new states. However, in the ratio of $\lb$ decays
to $D_s(2317)$ and $D_s(2460)$ relative to the $\bdbar$ decays to
these states the decay constants of $D_s(2317)$ and $D_s(2460)$
cancel, leading to robust predictions that can shed additional light
on the nature of these new states.

\subsection{Masses and Form Factors}

This assertion  that the diquark  in $\lb$ belongs in  the ${\overline
  3}$  under  color $SU(3)_c$  as  does  the  spectator anti-quark  in
  $\bdbar$ leads  to similar color interactions  has been dramatically
  confirmed  by relations  between  hadron masses  based  on a  simple
  QCD-based argument which goes beyond simple models for the spectator
  diquarks and the spectator anti-quark\cite{NewPenta,OlPenta}.

The hadrons  under consideration  all consist of  a quark,  denoted by
$q_i$, of any  flavor $i$ and ``light quark  brown muck" having either
the quantum numbers of a ${\overline 3}$ color diquark denoted by $ud$
or a  ${\overline 3}$ color anti-quark  denoted by $\bar  u$. While we
use the notations $ud$ and $\bar u$ for these light quark states, they
can apply to any more complicated light quark configuration containing
the same quantum numbers.

Consider the following four states of a quark of flavor $i$ bound to a
$ud$ or $\bar u$ configuration.  These are the pseudoscalar and vector
mesons
\beq 
\ket {P_i} = \ket {q_i \bar u}_{S=0}; ~ ~ ~ ~ ~ \ket {V_i} = \ket
{q_i \bar u}_{S=1}
\label{pi} 
\end{equation}
and  the isoscalar  and  isovector  baryons with  spins  1/2 and  3/2,
respectively
\beq 
\ket  {B^o_i} =  \ket {q_i  (ud)_{I=0}}_{S=1/2}; ~ ~  ~ ~  ~ \ket
{B^1_i} = \ket {q_i (ud)_{I=1}}_{S=3/2}
\label{boi} 
\end{equation}

Interesting    mass    relations    between   these    hadrons    were
obtained\cite{NewPenta} from the following QCD-motivated assumptions:
\begin{enumerate}
\item The effective mass of any constituent in a hadron depends on the
hadron wave  function only via  the color--electric field seen  by the
constituent.  The  color--electric fields  are very simply  related in
these hadrons.
\item The color--electric  field seen by the light  quark systems $ud$
and $\bar u$ are independent of the flavor of the quark $q_i$.
\item The color--electric field seen by the quark $q_i$ is independent
of  whether the color  ${\overline 3}$  light quark  system is  a $ud$
diquark or a $\bar u$ anti-quark.
\item The color--magnetic interaction  between the quark $q_i$ and the
spin-zero diquark vanishes in the baryon state $\ket {B^o_i}$.
\item The  color--magnetic contribution to the meson  mass cancels out
in the linear combination of masses\cite{NewPenta}
\beq 
\tilde M_i = {{3 M(V_i)+ M(P_i) } \over 4}
\label{pvi} 
\end{equation} 
\item The  hyperfine splitting between  the meson states  $\ket {P_i}$
and  $\ket {V_i}$  is  inversely proportional  to  the effective  mass
$m^{eff}_i$ of the quark of flavor $i$ and similarly for the hyperfine
splitting  between   the  baryon  states  $\ket   {B^o_i}$  and  $\ket
{B^1_i}$.   However   this  cancels   out   in   the  combination   of
Eq.~\ref{pvi}.
\end{enumerate} 
These immediately give for any two quark flavors i and j, 
\beq 
\tilde M_i - \tilde M_j = M(B^o_i) - M(B^o_j) \equiv m^{eff}_i - m^{eff}_j
\label{mij}
\end{equation}
and
\beq 
{{M(V_i) - M(P_i)}\over{M(V_j) - M(P_j)}} = {{M(B^1_i) -M(B^o_i)}\over{ M(B^1_j) - M(B^o_j)}} \equiv
{{m^{eff}_j}\over{m^{eff}_i}}
\label{hypij}
\end{equation} 

Eq.~\ref{mij} and Eq.~\ref{hypij} give all the mass relations between
mesons and baryons previously obtained
\cite{NewPenta,ICHJLmass,HJLMASS,Protvino,PBIGSKY} from the
Sakharov-Zeldovich model\cite{SakhZel} improved by DeRujula, Georgi
and Glashow \cite{DGG}. In particular we note that the change in
baryon masses when the $b$ quark in a $\Lambda_b$ is changed into a
$c$ quark to make a $\Lambda_c$,
\beq 
\langle m^{eff}_b-m^{eff}_c \rangle_{bar}= M(\Lambda_b)-M(\Lambda_c) =3339 \,{\rm MeV}
\label{barbc}
\end{equation} 
is exactly equal to the change in meson masses when the $b$ quark in a
$B$ meson is changed into a $c$ quark to make a $D$ when the
appropriate average of pseudoscalar and vector mesons is taken to
cancel out the hyperfine interaction.
\beq 
\langle
m^{eff}_b-m^{eff}_c \rangle_{mes}={{3(M_{B^{\scriptstyle*}}-M_{D^{\scriptstyle *}})+M_B-M_D}\over 4} =3342 \,{\rm MeV}. 
\label{mesbc}
\end{equation} 

The fact  that the  change in  the hadron mass  produced by  the quark
transition $b \rightarrow c$ is the  same when the quark is bound to a
$ud$ diquark  and to a $\bar  u$ anti-quark suggests  that the diquark
and anti-quark are  spectators in the transition and  will also effect
the transition $b  \rightarrow c$ in the same way  when it is produced
by the emission of a $W$ in a weak decay.

We now note that rearranging Eq. (\ref{barbc}) 
gives the dimensional constant $\bar{\Lambda}$ 
defined in \cite{georgi} to represent the effective mass of the light degrees of 
freedom in the $\lb$ and $\lc$ baryon 
\beq 
\bar{\Lambda} = m_{\Lambda_b}-m_b = m_{\Lambda_c}-m_c   
\label{lambdabarbc}
\end{equation} 
The value $\bar{\Lambda}= 575$ MeV
was estimated in Ref. \cite{NewPenta} using quark masses that fit both meson and
baryon masses. 

\subsection{Nonfactorization}

Although we will use a factorization assumption there is a question of
the correctness of such an assumption and what corrections would enter
from nonfactorization.  Nonfactorizable effects are known to be
important for hyperon and charmed-baryon nonleptonic decays \cite{SPT,
garcia, charmed}, 
An unambiguous signal for the presence of nonfactorizable effects in
$\lb$ decays
would be the observation of the decay $\Lambda_b \to \Sigma_c P$ or
$\Lambda_b \to \Sigma_c V$.  This is because, for the factorizable
contribution, the light diquark in the $\Lambda_b$ baryon remains
inert during the weak decay.  Thus, since the light diquark is an
isosinglet, and since strong interactions conserve isospin to a very
good approximation, the above $\Lambda_b$ decays are forbidden within
the factorization assumption\cite{dlo1}.

One way to estimate the  size of nonfactorizable corrections is to use
 the pole model.  In this  model, one assumes that the nonfactorizable
 decay  amplitude receives  contributions primarily  from one-particle
 intermediate  states, and that  these contributions  then show  up as
 simple poles in the decay  amplitude. Estimates of such pole diagrams
 in  $\Lambda_b$  decays  have been  found  to  be  small and  so  are
 neglected in  our analysis  \cite{lambdabold}.  Note that  these pole
 diagrams arise only through weak interactions involving the spectator
 quark  and  so small  estimates  of  the  pole diagram  confirms  the
 assumption   of  small  spectator   interaction  in   $\lb  (\bdbar)$
 decays\cite{dlo2}.

In the  next section we  discuss $\lb \to  \lc P$ decays while  in the
following  section we  discuss  $\lb  \to \lc  V$  decays. Finally  we
present our summary.

\section{Color allowed $\Lambda_b$ decays}
\subsection{$\Lambda_b \to \Lambda_c P$}

We begin our analysis by studying the nonleptonic decay $\Lambda_b \to
\Lambda_c P$.  The  general form for this amplitude  can be written as
\beq 
{\cal M}_P= A(\Lambda_b\to  \Lambda_c P) = i {\bar u}_{\Lambda_c}
(a + b\gamma_{5}) u_{\Lambda_b} ~.
\label{pscalar}
\eeq 
In the rest frame of the  parent baryon, the decay
amplitude reduces to 
\beq 
A(\Lambda_b\to \Lambda_c P) = i \chi_{\Lambda_c}^{\dagger} (S+P {\vec\sigma} \cdot {\hat p})
\chi_{\Lambda_b} ~,
\label{pscalar1}
\eeq 
where ${\hat p}$ is the unit vector along the direction of the
daughter baryon momentum, and the $S$ and $P$ wave amplitudes are
given by $S=\sqrt{2m_{\Lambda_b}(E_{\Lambda_c}+m_{\Lambda_c})}a$ and
$P=-\sqrt{2m_{\Lambda_b}(E_{\Lambda_c}-m_{\Lambda_c})}b$, where
$E_{\Lambda_c}$ and $m_{\Lambda_c}$ are, respectively, the energy and
mass of the final-state baryon $\Lambda_c$. The decay rate is then
given by
\beq 
\Gamma=\frac{{|\vec p|}}{8\pi{m_{\Lambda_b}}^{2}}(|S|^2
+|P|^2) ~,~~
\label{decayscalar}
\eeq 
where $|\vec{p}|$ is the magnitude of the momentum of the decay
products in the rest frame of the $\Lambda_b$.

We will use factorization in order to estimate various nonleptonic
amplitudes. The starting point is the SM effective Hamiltonian for
hadronic $B$ decays \cite{BuraseffH}:
\beq 
H_{eff}^q = {G_F \over \protect \sqrt{2}} [V_{ub}V^*_{uq}(c_1O_{1}^q + c_2 O_{2}^q) -
\sum_{i=3}^{10} V_{tb}V^*_{tq} c_i^t O_i^q] + h.c.,
\label{H_eff}
\eeq 
where 
\bea 
O_1^q  =  {\bar  q}_\alpha  \gamma_\mu  L c_\beta  \,  {\bar  c}_\beta
\gamma^\mu L b_\alpha &,& 
O_2^q = {\bar q} \gamma_\mu L c \, {\bar c} \gamma^\mu L b ~, \nn\\ 
O_{3(5)}^q = {\bar q} \gamma_\mu L b \, \sum_{q'} {\bar q}' \gamma^\mu
L (R) q' &,& 
O_{4(6)}^q = {\bar q}_\alpha \gamma_\mu L b_\beta \, \sum_{q'} {\bar q}'_\beta \gamma^\mu
L (R) q'_\alpha ~, \\ 
O_{7(9)}^q =  {3\over 2} {\bar q}  \gamma_\mu L b  \, \sum_{q'} e_{q'}
{\bar q}' \gamma^\mu R (L) q' &,& 
O_{8(10)}^q = {3\over 2} {\bar q}_\alpha \gamma_\mu L b_\beta \, \sum_{q'} e_{q'}
{\bar q}'_\beta \gamma^\mu R (L) q'_\alpha ~. \nn
\label{H_effops}
\eea 
In the above, $q$ can be either a $d$ or an $s$ quark, depending on
whether the decay is a $\Delta S = 0$ or a $\Delta S = -1$ process,
$q' = d$, $u$, $s$ or $c$, with $e_{q'}$ the corresponding electric
charge, and $R(L)= 1 \pm \gamma_5$.  The values of the Wilson
coefficients $c_i$ can be found in Ref. \cite{FSHe}:

We now apply the effective Hamiltonian to specific exclusive
$\Lambda_b$ and $B$ decays. We will focus on those processes for which
factorization is expected to be a good approximation, namely
color--allowed decays.

We begin with $\Lambda_b \to \Lambda_c \pi^-$ and $\overline{B}^0 \to
D^+ \pi^-$ which is a $b \to c \bar{u} d$ transition.  Factorization
allows us to write
\bea 
A(\Lambda_b\to \Lambda_c \pi^-)&=& if_{P} q^{\mu} \bra{\Lambda_c} \bar{c}
\gamma_\mu(1-\gamma_5)       b\ket{\Lambda_b}       X_{\pi}\nonumber\\
A(\overline{B}^0\to D^+  \pi^-)&=& if_{\pi} q^{\mu}  \bra{D^+} \bar{c}
\gamma_\mu(1-\gamma_5)     b\ket{\overline{B}^0}    X_{\pi}\nonumber\\
X_{\pi} & = &\frac{G_F}{\sqrt{2}} V_{cb}V_{ud}^* a_2 \
\label{amp-pi}
\eea 
The pseudoscalar decay constant $f_{\pi}$ is defined as 
\beq 
if_{\pi} q^{\mu}= \bra{\pi} \bar{d} \gamma^{\mu}(1 - \gamma_5)u \ket{0}
~, 
\eeq 
and $a_2=c_2+ c_{1} / {N_c}$.

Now,  the   vector  and  axial-vector  matrix   elements  between  the
$\Lambda_b$ and $\Lambda_c$ baryons can be written in the general form
\begin{eqnarray}
\bra{\Lambda_c} \bar{c} \gamma^\mu b \ket{\Lambda_b} & =& \bar{u}_{\Lambda_c} \left[ f_1 \gamma^\mu 
+ i \frac{f_2}{m_{\Lambda_b}}\sigma^{\mu\nu} q_\nu + \frac{f_3}{m_{\Lambda_b}} q^\mu \right] u_{\Lambda_b} \nonumber \\
\bra{\Lambda_c}  \bar{c} \gamma^\mu  \gamma_5 b  \ket{\Lambda_b}  & =&
\bar{u}_{\Lambda_c} \left[ g_1 \gamma^\mu 
+ i \frac{g_2}{m_{\Lambda_b}}\sigma^{\mu\nu} q_\nu + \frac{g_3}{m_{\Lambda_b}} q^\mu \right] \gamma_5 u_{\Lambda_b} ~,
\label{genamps}
\end{eqnarray}
where  the  $f_i$  and   $g_i$  are  Lorentz-invariant  form  factors.
Heavy-quark symmetry imposes constraints on these form factors. In our
approach  we  will  only  consider  the  $b$  as  heavy  and  consider
corrections up to order ${1 / m_c}$.  In the $m_b\to \infty$ limit
(but  with  $  {1  /  m_c}$ corrections),  one  obtains  the  relations
\cite{georgi} 
\bea 
f_1 &=& g_1= \left[1+\frac{\bar{\Lambda}}{2m_{\lc}}
(1-\frac{\bar{\Lambda}}{m_{\lc}}) 
\frac{\omega}{(\omega +1)}\right] \xi_B(\omega)+\frac{\eta(\omega)}{2m_c} \nonumber\\ 
f_2 & =& g_2 = f_3 = g_3 =-\frac{\bar{\Lambda}}{2m_{\lc}(\omega +1)} 
(1-\frac{\bar{\Lambda}}{m_{\lc}}) \xi_B(\omega)
\label{hqet}
\eea 
where $\xi_B(\omega)$ is the Isgur-Wise function for $\lb \to \lc$
transition,  $\bar{\Lambda}$ is defined in Eq.(\ref {lambdabarbc}),
$\eta(\omega)$ represents the correction from the kinetic energy of
heavy quark in the baryon and
$$\omega = {m_{\lb}^2+m_{\lc}^2 -q^2 \over 2m_{\lb}m_{\lc}}.$$
We point out that 
it is not necessary to estimate the quantity $\eta(\omega)$ 
for our calculation as we only use the relation $f_1=g_1$.
Estimates of $\eta(\omega)$ are found to be negligible\cite{korner}
and so we will set $\eta(\omega)=0$.    

The dimensional constant $\bar{\Lambda}$ representing
the effective mass of the light degrees of freedom in the $\lb$ and
$\lc$ baryon is estimated from Ref. \cite{NewPenta}
with $\bar{\Lambda}= 575$ MeV. 
Now from
Eq.~\ref{hqet} we see that in the $m_c \to \infty$ limit only the form
factors $f_1$ and $g_1$ are non zero and the form factors $f_2$,
$g_2$, $f_3$ and $g_3$ are suppressed by $O( 1/m_c)$. We will use this
fact later on in our calculations.
Using Eq.~\ref{amp-pi} and Eq.~\ref{hqet},  the amplitudes $a$ and $b$
of Eq.~\ref{pscalar} can be written as 
\bea 
a_{\pi} & = & f_{\pi} X_{\pi}\left[(m_{\Lambda_b}-m_{\Lambda_c}) f_1(q^2=m_{\pi}^2)
 +f_3\frac{m_{\pi}^2}{m_{\Lambda_b}} \right] ~, \nonumber\\ 
b_{\pi} & = & f_{\pi} X_{\pi}\left[(m_{\Lambda_b}+m_{\Lambda_c})
 g_1(q^2=m_{\pi}^2) -g_3\frac{m_{\pi}^2}{m_{\Lambda_b}} \right] ~.
\label{ab}
\eea 
In Eq.~\ref{ab} we can drop the suppressed contributions from the
$f_3(g_3)$  form factors  and using  the HQET  relation  $f_1=g_1$ the
quantities $a_{\pi}$ and  $b_{\pi}$ can be expressed in  terms of only
one form factor.  The $S$ and $P$ wave amplitudes are then written as,
\bea 
S &= &f_{\pi} X_{\pi}\left[(m_{\Lambda_b}^2-m_{\Lambda_c}^2)\right]
\sqrt{1-\frac{m_{\pi}^2}{(M_{\Lambda_b}+M_{\Lambda_c})^2}}
f_1(q^2=m_{\pi}^2)\nonumber\\
P &= &f_{\pi} X_{\pi}\left[(m_{\Lambda_b}^2-m_{\Lambda_c}^2)\right]
\sqrt{1-\frac{m_{\pi}^2}{(M_{\Lambda_b}-M_{\Lambda_c})^2}} f_1(q^2=m_{\pi}^2)\
\label{lambda-pi}
\eea

The   vector   and    axial-vector   matrix   elements   between   the
$\overline{B}^0$  and $D^+$  mesons can  be written  in terms  of form
factors \cite{BSW} 
\bea 
<D^+(p_D)|J_{\mu} | \bdbar > &=& \left[ (p_B+ p_D)_\mu - \frac{m_B^2-m_D^2}{q^2} q_\mu \right] F_1(q^2)\nonumber\\
&+ & \frac{m_B^2-m_D^2}{q^2} q_\mu F_0(q^2) \ 
\eea 
where $q = p_B -p_D$.  
{}From Eq.~\ref{amp-pi} one then obtains 
\bea A(\bdbar \to D^+ \pi^-)&=& f_{\pi}X_{\pi}(m_B^2-m_D^2) F_0(q^2=m_{\pi}^2)\
\label{B-pi}
\eea 
Note that the form of this amplitude is similar to the one in
Eq.~\ref{lambda-pi} with the important difference that for the $\lb$
decays there are two partial waves allowed by angular momentum
conservation.

We are interested here in the ratio 
\bea 
R_{\pi} &=& \frac{BR[\Lambda_b \to \Lambda_c \pi^-]} {BR[\bdbar \to D^+ \pi^-]}\
\label{P}
\eea 
We  can  define  similar  ratios  $R_K$,  $R_{D_s}$,  $R_D$  and
$R_{D_s(2317)}$. In passing we note that it is useful to also consider 
ratios of nonleptonic to semileptonic decays
\bea
SL_{\lb} & = & \frac{\Gamma[\lb \to \lc M]}
                    {d\Gamma[\lb \to \lc l \nu]/d\omega} \nonumber\\
SL_{\bdbar} & = & \frac{\Gamma[\bdbar \to D M]}
                    {d\Gamma [\bdbar \to D l \nu] /d \omega} \nonumber\\
SL_{\lb \bdbar} & = & \frac{{d\Gamma[\lb \to \lc l \nu]}{/d\omega}}
                    {d \Gamma [\bdbar \to D l \nu] /d \omega}\
\label{sl}
\eea
where $M$ is a $P$ or a $V$ meson. The  semileptonic
$ \lb \to \lc l \nu$ decay distribution \cite{korner}
as well as the nonleptonic $ \lb \to \lc M$ transition in factorization
 can be expressed in 
terms of the $ \lb \to \lc $ form factors in Eq.~\ref{genamps}. Now 
using Eq.~\ref{hqet} and the estimate of $\bar{\Lambda}$ the quantity
$SL_{\lb}$ is independent of form factors and can therefore be used to check 
for the validity of factorization in $\lb \to \lc M$ transitions.
One can use the ratio $SL_{\bdbar}$ to check for factorization in
$\bdbar$ decays. However the structure of the $1/m_{c,b}$ corrections are
not so simple here \cite{luke}. Finally the ratio $SL_{\lb \bdbar}$ can be 
used to express the ratio of $\lb \to \lc$ form factor and $\bdbar \to D$
form factor as a function of $\omega$.

For  the decays  $\lb \to  \lc  ( \pi^-,  K^-)$ there  are no  penguin
contributions.  However,  for the  decays $\lb \to  \lc (  D_s^-, D^-,
D_s(2317))$ there are penguin  contributions and the penguin operators
affect the $\lb$ and  $B$ decays differently\cite{bdl1}. For the decay
$\lb \to \lc D_s^-$ we obtain 
\beq 
A(\Lambda_b\to \lc D_s^-)= if_{D_s} q^{\mu} \bra{\lc} \bar{c} \gamma_\mu(1-\gamma_5) b\ket{\Lambda_b} X_K
+   i    f_{D_s}   q^{\mu}   \bra{\lc}\bar{c}   \gamma_\mu(1+\gamma_5)
b\ket{\Lambda_b} Y_K ~.
\label{ampDs}
\eeq
where 
\bea 
X_{D_s} & = & \frac{G_F}{\sqrt{2}} \left[ V_{cb}V_{cs}^* a_2- \sum_{q=u,c,t}V_{qb}V_{qs}^* (a_4^q +a_{10}^q)
\right] ~, \nonumber\\ 
Y_{D_s} & = & -\frac{G_F}{\sqrt{2}} \left[ \sum_{q=u,c,t}V_{qb}V_{qs}^* (a_6^q +a_8^q) \right] \chi_{D_s} ~,
\label{XY}
\eea 
with 
\beq 
\chi_{D_s} = \frac{2 m_{D_s}^2}{(m_s+m_c)(m_b-m_c)} ~
\label{XY1}
\eeq 
and for even ``i'', $a_i=a_i+ a_{i-1}/N_c$.
  
In the above equations we have used 
\beq 
i  f_{D_s}  q^{\mu}=  \bra{D_s}  \bar{s} \gamma^{\mu}(1  -  \gamma_5)c
\ket{0} ~, 
\eeq 
where $q^\mu \equiv p^\mu_{\lb} - p^\mu_{\lc} = p^\mu_{D_s}$ is the
four-momentum transfer. One can then show that
\beq 
\bra{D_s^-}  {\bar  s} (1  \pm  \gamma_5)  c  \ket{0} =  \mp  {f_{D_s}
  m_{D_s}^2 \over m_s +m_c} ~~,~~ 
\bra{\lc} {\bar  c} (1 \pm  \gamma_5) b \ket{\Lambda_b}  = {q^\mu\over
(m_b-m_c)} \bra{\lc} {\bar c} \gamma_\mu (1 \mp \gamma_5) b \ket{\Lambda_b} ~. 
\label{LRtrans}
\eeq 
This then leads to 
\bea 
a_{D_s} & = & f_{D_s} (X_{D_s}+Y_{D_s})\left[(m_{\lb}-m_{\lc})f_1
+f_3\frac{m_{D_s}^2}{m_{\Lambda_b}} \right] ~, \nonumber\\ 
b_{D_s} & =& f_{D_s} (X_{D_s}-Y_{D_s})\left[(m_{\Lambda_b}+m_{\lc})g_1
-g_3\frac{m_{D_s}^2}{m_{\Lambda_b}} \right] ~.  
\eea 
and 
\bea S &= &f_{D_s} (X_{D_s}+Y_{D_s}) \left[(m_{\Lambda_b}^2-m_{\Lambda_c}^2)\right]
\sqrt{1-\frac{m_{D_s}^2}{(M_{\Lambda_b}+M_{\Lambda_c})^2}}
f_1(q^2=m_{D_s}^2)\nonumber\\ 
P &= &f_{D_s}( X_{D_s} - Y_{D_s}) \left[(m_{\Lambda_b}^2-m_{\Lambda_c}^2)\right]
\sqrt{1-\frac{m_{D_s}^2}{(M_{\Lambda_b}-M_{\Lambda_c})^2}} f_1(q^2=m_{D_s}^2).\
\label{lambda-Ds}
\eea 
The corresponding $B$ decay, $ \bdbar \to D^+ D_s^-$, is 
\bea
A(\bdbar \to D^+ D_s^-)&=& f_{D_s}(X_{D_s}+Y_{D_s})(m_B^2-m_D^2 F_0(q^2=m_{D_s}^2)\
\label{B-Ds}
\eea 
Similar  expressions can be written  for the pair  of decays $\lb
\to \lc D^-$ and $\bdbar \to D^+ D^-$ with obvious changes.  Note that
from Eq.~\ref{XY} and Eq.~\ref{XY1} the quantity $Y_{D_s}$ or 
$\chi_{D_s}$ is formally
suppressed by ${ 1/ m_b}$ though with a large coefficient.
Taking the effective quark masses $m_b=5.050$ GeV,
$m_c=1.710$ GeV and $m_s=0.602$ GeV  \cite{NewPenta} we find
$\chi_{D_s} \sim $ 1, which shows the effect of the large coefficient. 
However, to
 simplify our discussion we will neglect
$Y_{D_s}$.  Given the fact that the penguins are smaller than the tree
amplitude, the error from the neglect of $Y_{D_s}$ is of the same
order as the sub-leading ${ 1/ m_b}$ effects which we have
neglected.  We should point out that for CP violating studies the
quantities $X_{D_s}$ and $Y_{D_s}$ play an important role \cite{bdl1}. 
However here we are interested in decay rates only and not CP--violating observables.

Using the  values of the particle  masses as well as  the lifetimes of
the $\Lambda_b$ and $\bd$ \cite{PDG} we obtain 
\bea 
R_{\pi} & =& 1.73\frac{f_1^2( q^2=m_{\pi}^2)}{F_0^2(q^2=m_{\pi}^2)}\
\label{e-input}
\eea 
Now  using  Eq.~\ref{e-input}  and experimental  information  on
$R_{\pi}$   allows    us   to   extract   the    form   factor   ratio
$$r(q^2=m_{\pi}^2)=f_1^2(q^2=m_{\pi}^2) / {F_0(q^2=m_{\pi}^2)}.$$  

There has been a preliminary measurement of $\lb \to \lc \pi^-$ by
CDF\cite{CDF} with the branching ratio $ (6.0 \pm 1.0(stat) \pm 0.8
(syst) \pm 2.1(BR)) \times 10^{-3}$.  Using the PDG value for $\bdbar
\to D^+ \pi^-$ which is $(2.76 \pm 0.25) \times 10^{-3}$ \cite{PDG}
and taking the central value of the measurements we obtain $R_{\pi}
\approx 2.17$.  This then leads to, using Eq.~\ref{e-input} 
\bea
\frac{f_1( q^2=m_{\pi}^2)}{F_0(q^2=m_{\pi}^2)} & = & 1.12.\ 
\label{cdfnum}
\eea 
In the heavy $m_c$ and $m_b$ limit we can relate the form factors
$f_1$ and $F_0$  to the Isgur-Wise functions for $\lb  \to \lc$ and $B
\to D$ transition, $\xi_B(\omega_B)$ and $\xi_M(\omega_M)$ 
\bea f_1(m_{\pi}^2) & \approx & f_1(0)= \xi_B(\omega_B^{max}) \nonumber\\
F_0(m_{\pi}^2) & \approx & F_0(0) = F_1(0)=\frac{m_B+m_D}{2 \sqrt{m_Bm_D}}\xi_M(\omega_M^{max})\
\label{FF0}
\eea 
 This then leads to 
\bea 
\xi_B(\omega_B^{max}) & = & 1.4 \xi_M(\omega_M^{max}) \
\label{isgurwise}
\eea 
In the heavy $m_c$ and $m_b$ limit $\omega_B= \omega_M$. However for
actual masses $\omega_B^{max}=1.458$ and $\omega_M^{max}=1.588$ which
indicates that $m_c \to \infty$ is not a very good limit.  Keeping in
mind that $\xi_{B,M}(\omega=1)=1$, Eq.~\ref{isgurwise} indicates that
the baryon Isgur-Wise function falls off slower than the mesonic
counterpart.

To make predictions for the ratio  $R_P$ for the other decays we would
need  the ratio of  form factors  $r(q^2=m_{P}^2)=f_1^2(q^2=m_{P}^2) /
{F_0(q^2=m_{P}^2)}$. This requires a dynamical input which will be our
only   assumption  for   the   calculation  of   the  decays   besides
factorization.

We assume  a general parameterization of the  form factors for the
region of $q^2$ that we are interested in 
\bea 
f_1(q^2) &=& f_1(0)\eta_B({q^2\over M_{B^*}^{2}})\nonumber\\ 
F_0(q^2)&=& F_0(0) \eta_M({q^2\over M_{M^*}^{2}})\nonumber
\label{ffactor}
\eea 
where $M_{B^*}$ and $M_{M^*}$ are some heavy masses that scale as
$m_b$.  In  other words  the difference $M_{B^*}-M_{M^*}$  vanishes as
$m_b \to  \infty$.  Furthermore $\eta_{B,M}(0)=1$  by definition. 
Assuming $q^2$ to  be smaller than $M_{B^*}^2$ and  $M_{M^*}^2$ we can
write 
\bea 
\eta_B({q^2  \over   M_{B^*}^{2}})  &   =  &  1   +  \alpha_B{q^2\over
  M_{B^*}^{2}}+..  \nonumber\\ 
\eta_M({q^2 \over M_{B^*}^{2}}) & = & 1 + \alpha_M{q^2\over M_{B^*}^{2}}+.. 
\label{exp}
\eea 
and so 
\bea 
\frac{\eta_B({q^2    \over     M_{B^*}^{2}})}    {\eta_M({q^2    \over
    M_{B^*}^{2}})} & = & 1 + \alpha_B{q^2\over
M_{B^*}^{2}}-\alpha_M{q^2\over M_{B^*}^{2}}+..\
\label{ratio}
\eea 
Note that the often used pole model for form factors is just one
example of the general parameterization in Eq.~\ref{ffactor} where
$M_{B^*}$ and $M_{M^*}$ can be identified with the excited baryon and
meson states

Now the  largest $q^2$ we will  be interested in is $  q^2 \sim 4
GeV^2$ and so taking $M_{M^*,B^*}$ around 5-6 GeV we expect the second
term in Eq.~\ref{exp}  to be around 10-15 \%.   Furthermore, we expect
$\alpha_B$ and $\alpha_M$  to be of the same sign  as the form factors
increase with $q^2$.  This  implies further cancellation in the second
term in Eq.~\ref{ratio}.  and so to a good approximation 
\bea
\frac{\eta_B({q^2\over M_{B^*}^{2}})} {\eta_M({q^2\over M_{M^*}^{2}})}
& = & 1\ 
\eea 
So in the heavy $m_b$ limit we can write for the form factor ratio $r$
\bea r(q^2) & \approx & r(q^2=0) \approx r(q^2=m_{\pi}^2) 
\eea 
Hence the measurement
of $r(q^2=m_{\pi}^2)$  allows us to make predictions  for other decays
which are presented in in Table.~\ref{pseudoscalar}.
\begin{table}
\caption{Table for $R_P=\frac{BR[\Lambda_b \to \Lambda_c P^-]}
             {BR[\bdbar \to D^+ P^-]}$ with experimental input}
\begin{center}
\begin{tabular}{|c|c|c|}
\hline 
$R_P$ & Theory & Experiment \\ 
\hline 
$R_{\pi}$ & $2.17$ & $2.17 \cite{CDF}$ \\ 
\hline 
$R_{K}$ & $2.14$ & $- -$ \\ 
\hline 
$R_{D}$ & $1.79$ & $- -$ \\ 
\hline 
$R_{D_s}$ & $1.75$ & $- -$ \\ 
\hline
$R_{D_s(2317)}$ & $1.58$ & $- -$ \\ 
\hline
\end{tabular}
\end{center}
\label{pseudoscalar}
\end{table}

\subsection{$\Lambda_b \to \Lambda_c V$}
We now turn  to the decays $\Lambda_b\to \Lambda_c  V$ where $V= \rho,
K^* a_1,  D^*, D_s^*,  D_s(2460)$ The general  decay amplitude  can be
written as \cite{SPT, bdl1} 
\beq 
{\cal M}_V= Amp(\Lambda_b\to \Lambda_c V) = {\bar u}_{\Lambda_a} \varepsilon^*_\mu \left[
\frac{p_{\Lambda_b}^\mu     +    p_{\Lambda_c}^\mu}    {m_{\Lambda_b}}
(a+b\gamma_{5}) + \gamma^{\mu} (x+y\gamma_{5}) \right] u_{\Lambda_b}~,
\label{vector}
\eeq 
where  $\varepsilon^*_\mu$ is  the  polarization  of the  vector
meson.  In  the rest  frame of  the $\Lambda_b$, we  can write  $p_V =
(E_V,0,0,|\vec{p}|)$           and           $p_{\Lambda_c}          =
(E_{\Lambda_c},0,0,-|\vec{p}|)$ and Eq.~\ref{vector} can be reduced to
\cite{SPT} 
\bea 
{\cal M}_V & =& \chi^{\dagger}_f \left[S \vec{\sigma}+
P_1 \hat{p} +iP_2 \hat{p} \times {\vec{\sigma}}
+D(\vec{\sigma} \cdot \hat{p})\hat{p} \right]\cdot \vec{\epsilon}\chi_i\
\label{vector_restframe}
\eea 
where $\hat{p}$  is a unit vector in the  direction of the vector
meson momentum.  The  amplitudes for the three helicity  states of the
vector meson can be written as 
\bea 
{\cal M}(+1) & = & \frac{P_2-S}{\sqrt{2}} \chi_f^{\dagger}\left[\vec{\sigma}.(\vec{\epsilon_1}+i\vec{\epsilon_2})
\right]\chi_i\nonumber\\ 
{\cal M}(-1) & = & \frac{P_2+S}{\sqrt{2}} \chi_f^{\dagger}\left[\vec{\sigma}.(\vec{\epsilon_1}-i\vec{\epsilon_2})
\right]\chi_i\nonumber\\ 
{\cal M}(0) & = & \frac{E_V}{m_V }
\chi_f^{\dagger}\left[(S+D)\vec{\sigma} \cdot \vec{p} +P_1 \right]\chi_i\
\label{helicity}
\eea 
In  terms of the  quantities defined in Eq.~\ref{vector}  we then
have 
\bea 
S    &   =    &   -\sqrt{2m_{\Lambda_b}(E_{\Lambda_c}+m_{\Lambda_c})}y
\nonumber\\ 
P_1     &     =    &\sqrt{2m_{\Lambda_b}(E_{\Lambda_c}+m_{\Lambda_c})}
\frac{p}{E_V}\left[\frac{m_{\Lambda_b}+m_{\Lambda_c}} {E_{\Lambda_c} +
m_{\Lambda_c}}x + 2a \right] \nonumber\\ 
P_2   &    =   &   -\sqrt{2m_{\Lambda_b}(E_{\Lambda_c}+m_{\Lambda_c})}
\frac{px}{E_{\Lambda_c} +
m_{\Lambda_c}}\nonumber\\ 
D & = & \sqrt{2m_{\Lambda_b}(E_{\Lambda_c}+m_{\Lambda_c})} \frac{p^2}{E_V(E_{\Lambda_c} +m_{\Lambda_c})}\left[2b-y \right]\ 
\eea
We note from Eq.~\ref{helicity} that for light $V$, $E_V \sim m_{\lb}$
and so as $m_b \to \infty$ the amplitude with longitudinally polarized
$V$ dominates.  Hence  in this limit only two  combinations of partial
waves contribute.  We also note that the longitudinal amplitude ${\cal
M}(0)$ is  of the  same form  as Eq.~\ref{pscalar1} for  $ \lb  \to \lc
P$. Hence in the  $m_b \to \infty$ and a light $V$  limit we can write
the decay  rate for $\lb \to \lc  V$, following Eq.~\ref{decayscalar},
as 
\beq 
\Gamma_{V0}=\frac{{|\vec p|}}{8\pi{m_{\Lambda_b}}^{2}} \left[(|S+D|^2 +|P_1|^2)\frac{E_V^2}{m_V^2} \right] ~,~~
\label{decayvector}
\eeq 
The complete expression for  the decay rate with finite $m_b$ and
$V$ not necessarily light is given by 
\beq 
\Gamma_V=\frac{{|\vec p|}}{8\pi{m_{\Lambda_b}}^{2}} \left[(|S+D|^2
+|P|^2)\frac{E_V^2}{m_V^2} +(|S|^2+|P_2|^2) \right] ~,~~
\label{decayvectorfull}
\eeq 
where $|\vec{p}|$  is the magnitude of the  momentum of the decay
products in the rest frame of the $\Lambda_b$.

We use factorization  to calculate the coefficients $a$,  $b$, $x$ and
$y$ in Eq.~\ref{vector} for  various decays.  Consider first the decay
$\Lambda_b  \to  \Lambda_c  \rho$.    We  define  the  decay  constant
${g_{\rho}}$ as 
\beq 
m_{\rho}g_{\rho}\varepsilon_{\mu}^{*} = \bra{\rho} \bar{d}\gamma_{\mu}u\ket{0} ~. 
\label{rhovector}
\eeq 
and so we obtain 
\bea 
A(\Lambda_b\to \Lambda_c \rho) &=& m_{\rho} g_{\rho} \left\{ \varepsilon_{\mu}^{*} \bra{\Lambda_c} \bar{c}
\gamma^\mu(1-\gamma_5)   b\ket{\Lambda_b}   X_{\rho}  \right.    \nn\\
X_{\rho} &= &\frac{G_F}{\sqrt{2}}V_{cb}V_{ud}^* a_2 \ 
\eea 
with $a_2 = c_2 + c_1/N_c$, and $a$, $b$, $x$ and $y$ in Eq.~\ref{vector} given by 
\bea 
a_{\rho} &=& m_{\rho} g_{\rho}f_{2} X_{\rho}, \nn\\ 
b_{\rho} &=& -m_{\rho}g_{\rho}g_{2} X_{\rho} ~, \nn\\ 
x_{\rho} &=& m_{\rho} g_{\rho} [f_{1}- \frac{m_{\Lambda_c}+m_{\Lambda_b}}{m_{\Lambda_b}}f_{2}] X_{\rho} ~,
\nonumber\\ 
y_{\rho} &=& -m_{\rho} g_{\rho} [g_{1}+ \frac{m_{\Lambda_b}-m_{\Lambda_c}}{m_{\Lambda_b}}g_{2}] X_{\rho} ~. 
\label{abxyvec}
\end{eqnarray}
{}For the general decay $\lb  \to \lc V$ the quantities $a$, $b$, $x$
and $y$ have the same form  as Eq.~\ref{abxyvec} and we can then write
\bea 
S+D & = & 2\mv g_V \M \left[f_1 +f_2\frac{\mv^2}{(\M-\m)\M} \right] \frac{K_1}{K_2}\nonumber\\ 
P_1   &   =  &2\mv   g_V   \M  \left[f_1   -f_2\frac{\mv^2}{(\M+\m)\M}
\right]\frac{ \M^2-\m^2}{\M^2+\m^2}\frac{K_4}{K_3K_1}\nonumber\\ 
P_2 &  = & -\mv g_V
\M \left[f_1 -f_2 \right]\frac{\M-\m}{\M} \frac{K_4}{K_1}\nn\\ 
S & = &\mv g_V \M \left[f_1 +f_2 \right]\frac{\M+\m}{\M} K_1\
\label{SPD}
\eea 
where 
\bea 
K_1 & = & \sqrt{1-\frac{\mv^2}{(\M+\m)^2}}\nn\\ 
K_2 & = & 1+\frac{\mv^2}{(\M^2-\m^2)}\nn\\ 
K_3 & = & 1-\frac{\mv^2}{(\M^2+\m^2)}\nn\\ 
K_4 & =&\sqrt{1-\frac{2\mv^2(\M^2+\m^2)}{(\M^2-\m^2)^2}\left[1-\frac{\mv^2}{2(\M^2+\m^2)}
\right] }\ 
\eea 
In  the   light  $V$  and  $m_b  \to  \infty$  case
$K_{1,2,3,4}  \to 1$  and the  dependence on  the form  factor $f_2$
drops out.  Also, only the first two combination of
partial waves, $S+D$  and $P_1$ contribute.  In the  heavy $m_b$ limit
and identifying the light $V= \rho$, as an example, we can write, using the
relations   in  Eq.~\ref{hqet}  and   dropping  terms   suppressed  by
${m_{\rho}^2 / E_{\rho}^2}$, 
\bea |M|^2(\Lambda_b \to \Lambda_c \rho^-) & = & ({{G_F}\sqrt 2})^2 |V_{cb}V_{ud}^*|^2 a_2^2 m_{\rho}^2f_{\rho}^2
f_1(m_{\rho}^2)^2
m_{\Lambda_b}^2\frac{E_{\rho}^2}{m_{\rho}^2}\nonumber\\ 
& & \left[1+ \frac{(m_{\Lambda_b}^2-m_{\Lambda_c}^2)^2} {(m_{\Lambda_b}^2+m_{\Lambda_c}^2)}\right]\
\label{lambdabrho}
\eea 
The corresponding  expression for ${\bar B}^0 \to  D^+ \rho^-$ is
within the factorization assumption \cite{datta-vary}, 
\bea 
|M|^2( {\bar B}^0 \to D^+ \rho^-) & = & ({{G_F}\sqrt 2})^2 |V_{cb}V_{ud}^*|^2 a_2^2
m_{\rho}^2f_{\rho}^2 F_1(m_{\rho}^2)^2 m_B^2\frac{p^2}{m_{\rho}^2}\
\label{brho}
\eea 
{} From Eq.~\ref{brho} we see, that unlike the pseudoscalar case,
the form factor  $F_1(q^2)$ appears.  However, $F_0(q^2=0)=F_1(q^2=0)$
and for  the values  of $q^2$ we  are interested  in we will  make the
assumption $F_1(q^2)  \approx F_0(q^2)$.  We therefore  obtain for the
ratio of form factors
\bea
r(q^2) &= &\frac{f_1^2( q^2)}{F_0(q^2)} \approx \frac{f_1^2( q^2))}{F_1(q^2)} \approx r(q^2=0)\
\label{vectorinput}
\eea 
We  can now  use  the  experimental  input for  $r(q^2=m_{\pi}^2)$  from
Eq.~\ref{e-input} to make predictions for the various $ \lb \to \lc V$
decays.

It is clear from Eq.~\ref{SPD} that as $m_V$ gets larger the effect
of the form factor $f_2$ becomes important and we have to introduce
additional model dependency by requiring the value of $f_2$. However
$f_2$ is suppressed by $ { 1 / m_c}$ and so we will present our
predictions in two cases. In the first case we shall take the $m_c
\to \infty$ limit and so $f_2=0$.  However, we will use the measured
values of the various particle masses thereby including finite $m_c$
effects. Hence, the only assumption that we make here is that $m_c
\to \infty$ is applicable only as far as the form factor $f_2$ is
concerned. For the second case we estimate ${f_2 / f_1}$ using
Eq.~\ref{hqet} with $m_c=1.710$,  and
${\bar{ \Lambda}}=0.575$ GeV \cite{NewPenta} 
and $\xi_B(\omega) \approx 1$.

We present  our results in Table.~\ref{vectora} with  $f_2=0$ while in
Table.~\ref{vectorb} we present results with $f_2 \ne 0$.
\begin{table}
\caption{Table for $R_V=\frac{BR[\Lambda_b \to \Lambda_c V^-]}
             {BR[\bdbar \to D^+ V^-]}$ for $f_2=0$ }
\begin{center}
\begin{tabular}{|c|c|c|}
\hline 
$R_V$ & Theory($\Gamma_V$) & Theory($\Gamma_{V0}$) \\ 
\hline
$R_{\rho}$ & $1.75$ & $1.68$ \\ 
\hline 
$R_{K^*}$ & $1.82$ & $1.72$ \\
\hline 
$R_{a_1}$ & $2.08$ & $ 1.89$ \\ 
\hline 
$R_{D^*}$ & $3.21$ & $2.58$ \\ 
\hline 
$R_{D_{s}^*}$ & $3.47$ & $2.74$ \\ 
\hline
$R_{D_s(2460)}$ & $4.76$ & $3.50$ \\ 
\hline
\end{tabular}
\end{center}
\label{vectora}
\end{table}
The  second column  in  Table.~\ref{vectora} and  Table.~\ref{vectorb}
uses  the full  decay rate  in Eq.~\ref{decayvectorfull}  while column
three uses the  decay rate with only the  longitudinal polarization as
given in Eq.~\ref{decayvector}.   {} From Table.~\ref{vectora} we make
the following observations.  When the vector meson $V$ is light then there
is little difference between the entries in column two and column three
indicating   the    dominance   of   the    longitudinally   polarized
contribution. With higher $m_V$  the contributions from the transverse
polarization components become  important.  The second observations is
that,  for  light  $V$,  $R_V   \le  2$  as  only  two  partial  waves
corresponding  to  the  longitudinal vector  polarization  contribute.
However,  with increasing $m_V$  the various  quantities $K_{1,2,3,4}$
become important  and in particular the partial  wave $P_1$ increases.
The  net  effect is  that,  even  with  only the  longitudinal  vector
polarization, the $\lb$ decay rate  is more than the corresponding $B$
rate by more than a factor of two for charm final states. Finally we see that the branching
ratio for  $\lb \to \lc  D_s(2460)$ is between  four to 5  times times
that of the corresponding $B$ mode.  This is simply from the fact that
more partial  waves contribute in the  $\lb$ decays and  the fact that
the  $ \lb  \to  \lc$ form  factor  is larger  than the  corresponding
$\bdbar \to D^+$ form factor as suggested by experiment \cite{CDF}.
\begin{table}
\caption{Table for $R_V=\frac{BR[\Lambda_b \to \Lambda_c V^-]}
             {BR[\bdbar \to D^+ V^-]}$ for $f_2 \ne 0$ }
\begin{center}
\begin{tabular}{|c|c|c|}
\hline 
$R_V$ & Theory($\Gamma_V$) & Theory($\Gamma_{V0}$) \\ 
\hline
$R_{\rho}$ & $1.75$ & $1.68$ \\ 
\hline 
$R_{K^*}$ & $1.81$ & $1.72$ \\
\hline 
$R_{a_1}$ & $2.07$ & $1.88$ \\ 
\hline 
$R_{D^*}$ & $3.17$ & $2.56$ \\ 
\hline 
$R_{D_s^*}$ & $3.43$ & $2.71$ \\ 
\hline
$R_{D_s(2460)}$ & $4.68$ & $3.46$ \\ 
\hline
\end{tabular}
\end{center}
\label{vectorb}
\end{table}
{} From Table.~\ref{vectorb} we see that the effects of non zero $f_2$
from finite $m_c$ effects are rather small.
\section{Summary}
In  this   paper  we  have   considered  nonleptonic  Cabibbo--allowed
$\Lambda_b$  decays  in  the  factorization  approximation.   We  have
discussed possible nonfactorizable effects  and how experiments can be
used  to test  look for  them.   We calculated  decays of  the type  $
\Lambda_b \to \Lambda_c P $ and $ \Lambda_b \to \Lambda_c V $ relative
to $\bdbar \to  D^+ P$ and $\bdbar \to D^+ V$  where we included among
the  pseudoscalar  states(P)  and   the  vector  states(V)  the  newly
discovered  $D_s$  resonances, $D_s(2317)$  and  $D_s(2460)$. Using  a
preliminary measurement of the branching ratio for $\lb \to \lc \pi^-$
and a  mild assumptions  about the $q^2$  behavior of form  factors we
made   predictions  for   several   $\lb$  decays   relative  to   the
corresponding $\bdbar$  decays.  In general we found  the $\lb$ decays
to be larger than the  corresponding $\bdbar$ decays and in particular
we found  $\lb \to \lc  D_s(2460)$ can be  between four to  five times
$\bdbar  \to  D^+  D_s(2460)$.   This  enhancement  of  $\lb$  can  be
understood from the  fact that more partial waves  contribute in $\lb$
decays than  in $\bdbar$ decays and the  fact that the $  \lb \to \lc$
form factor  is larger  than the corresponding  $\bdbar \to  D^+$ form
factor.

{\bf Acknowledgments}:

We thank Shin-Shan Yu for useful discussions and bringing the CDF
measurements to our notice. This work was financially supported by
NSERC of Canada.


\end{document}